\documentclass[%
reprint,   
 amsmath,amssymb,
 aps,
]{revtex4-2}
\usepackage{enumitem}
\usepackage{graphicx}
\usepackage{dcolumn}
\usepackage{bm}
\usepackage{hyperref}
\usepackage{slashed}
\usepackage{subfloat}
\usepackage{multirow}
\usepackage[table]{xcolor}
\usepackage{pgfplots,subfigure}
\usepackage{orcidlink}
\usepackage{float}
\usepackage{geometry}
\usepackage[table]{xcolor} 
\usepackage{booktabs}       
\usepackage{array}          
\usepackage{hyperref}

\newcolumntype{C}{>{\centering\arraybackslash}m{4cm}} 
\definecolor{acsblue}{RGB}{17,76,139}

\begin{document}

\fontsize{7.6}{8.6}\selectfont
\title{Wave Propagation and Effective Refraction in Lorentz-Violating Wormhole Geometries}

\author{Semra Gurtas Dogan\orcidlink{0000-0001-7345-3287}}
\email{semragurtasdogan@hakkari.edu.tr}
\affiliation{Department of Medical Imaging Techniques, Hakkari University, 30000, Hakkari, Türkiye}

\author{Omar Mustafa\orcidlink{0000-0001-6664-3859}}
\email{omar.mustafa@emu.edu.tr (Corresponding Author)}
\affiliation{Department of Physics, Eastern Mediterranean University, 99628, G. Magusa, north Cyprus, Mersin 10 - Türkiye}

\author{Abdulkerim Karabulut\orcidlink{0000-0003-1694-5458}}
\email{akerimkara@gmail.com}
\affiliation{Department of Basic Sciences, Erzurum Technical University, 25050, Erzurum, Türkiye}

\author{Abdullah Guvendi\orcidlink{0000-0003-0564-9899}}
\email{abdullah.guvendi@erzurum.edu.tr}
\affiliation{Department of Basic Sciences, Erzurum Technical University, 25050, Erzurum, Türkiye}

\date{\today}

\begin{abstract}
{\fontsize{7.6}{8.6}\selectfont \setlength{\parindent}{0pt}  
We study the propagation of massless scalar waves in static, spherically symmetric Lorentz-violating wormhole spacetimes within a geometric-optical framework. Starting from a general metric characterized by an arbitrary lapse function and areal radius, we derive curvature invariants, establish regularity conditions at the wormhole throat, and reduce the Klein-Gordon equation to a Helmholtz-type radial wave equation. This formulation naturally leads to a position- and frequency-dependent effective refractive index determined by the underlying spacetime geometry and Lorentz-violating structure, resulting in effective frequency-dependent wave-optical behavior. We show that divergences of the refractive index coincide with Killing horizons, while curvature-induced turning points control reflection, transmission, and confinement of scalar waves. By analyzing constant, linear, and quadratic lapse profiles, we identify horizonless transmission regimes, asymmetric wave propagation, and multi-horizon trapping structures. Our results reveal that Lorentz violation can significantly modify wave-optical properties of curved spacetime, generating graded-index analogues and geometric confinement of modes without curvature singularities. This unified optical perspective provides a robust framework for investigating wave scattering, resonances, and potential observational signatures in Lorentz-violating gravitational backgrounds.}
\end{abstract}


\maketitle


\section{Introduction}
\label{sec:intro}

\setlength{\parindent}{0pt}

Traversable wormholes represent one of the most fascinating and conceptually challenging predictions of gravitational physics, offering spacetime geometries in which distant regions of the universe are connected through a nontrivial topology~\cite{TW1,TW2,TW3,TW4}. Since the seminal work of Morris and Thorne, it has been understood that such configurations cannot be supported within classical general relativity (GR) without violating the null energy condition (NEC), thereby requiring the presence of exotic matter at the throat~\cite{MorrisThorne1988}. This requirement poses a serious obstacle to the physical realization of wormholes, as it conflicts with the standard energy conditions obeyed by known classical matter fields. Consequently, a major line of research in recent decades has focused on identifying theoretical frameworks in which wormhole solutions can arise without invoking fundamentally pathological matter sources~\cite{TW5,TW6,TW7,TW8,TW9}.

\vspace{0.01cm}
\setlength{\parindent}{0pt}

Modified theories of gravity provide a natural setting in which this difficulty may be alleviated. In such theories, effective violations of energy conditions can emerge from geometric corrections, higher-curvature terms, or background-field contributions rather than from the matter sector itself~\cite{TW5,TW6,TW7,TW8,TW9,Radhakrishnan2024}. Lorentz invariance is a cornerstone symmetry underlying both GR and the Standard Model of particle physics~\cite{Kibble,Smolin,C-1,C-2}. It ensures the universality of the speed of light and determines the local causal structure of spacetime. Despite its extraordinary experimental success at accessible energy scales, Lorentz invariance is widely expected to be only an approximate symmetry that may be deformed, violated, or spontaneously broken in candidate theories of quantum gravity and high-energy effective field theories~\cite{Kibble,Smolin,C-1,C-2,C-3,C-4,C-5}. A systematic and model-independent framework for describing such effects is provided by the Standard-Model Extension (SME), in which Lorentz-violating operators arise from couplings between conventional fields and background tensor fields that acquire nonvanishing vacuum expectation values~\cite{KS1989,ColladayKostelecky1997,C-2,C-1}. These background fields introduce preferred spacetime directions, thereby modifying gravitational dynamics while preserving observer covariance and the effective field-theoretic structure. Systematic studies within the SME framework have placed stringent bounds on Lorentz-violating parameters across numerous sectors, including charge conjugation, parity, and time-reversal (CPT) violation~\cite{23,24,25,26}, radiative corrections~\cite{27,28,29,30,31,32,33,34}, CPT-even and CPT-odd gauge sectors~\cite{35,36,37,38,39}, photon-fermion couplings~\cite{40,41,42,43}, and fermion interactions~\cite{44,45,46,47}.

\vspace{0.01cm}
\setlength{\parindent}{0pt}

Within Lorentz-violating gravity, a growing body of work has demonstrated that wormhole geometries can be supported in a consistent and controlled manner. In Einstein-bumblebee gravity, for example, a vector field undergoes spontaneous Lorentz symmetry breaking, giving rise to exact traversable wormhole solutions in which the Lorentz-violating parameters determine both the throat geometry and the degree of effective NEC violation \cite{Ovgun2019}. Subsequent studies have explored the dynamical and perturbative properties of these solutions, including their quasi-normal mode spectra, providing evidence for their stability within the effective theory regime \cite{Oliveira2018}. Extensions incorporating additional matter content, such as phantom scalar fields, have further enlarged the solution space, showing how Lorentz symmetry breaking can act as a unifying mechanism to sustain nontrivial topologies \cite{Ding2024}. Beyond vector-field models, antisymmetric tensor fields offer a compelling and physically motivated generalization. In particular, the Kalb-Ramond field, which naturally appears in the low-energy limit of string theory, can induce spontaneous Lorentz symmetry breaking when treated as a background rank-2 tensor. Early investigations focused primarily on black hole solutions modified by Kalb-Ramond backgrounds, revealing distinctive imprints on horizon structure, thermodynamics, and geodesic motion \cite{Maluf2020, Ghosh2024}. More recently, it has been shown that such Lorentz-violating black hole geometries can be related to traversable wormholes through thin-shell constructions, establishing a direct and nontrivial connection between black hole and wormhole solutions in antisymmetric tensor gravity \cite{MalufMuniz2021}. These developments point toward a unifying picture in which Lorentz-violating gravity provides a consistent framework for treating black holes and wormholes on equal footing. In this perspective, background vector or tensor fields effectively source nontrivial spacetime geometries, relocating the violation of energy conditions from exotic matter to geometric or vacuum-field sectors~\cite{KS1989, C-1}. This shift not only improves theoretical consistency, but also opens new avenues for studying the physical implications of wormholes, including their causal structure, stability, and interaction with test fields \cite{LV}. In particular, the propagation of waves in such backgrounds offers a powerful probe of the underlying geometry, as it encodes both local curvature effects and global topological features \cite{abdelghani-2025,o1}.

\vspace{0.01cm}
\setlength{\parindent}{0pt}

The aim of this work is to develop a unified and geometrically transparent description of massless scalar-wave propagation in Lorentz-violating wormhole spacetimes, with special emphasis on how curvature alone can induce optical-like phenomena in vacuum. Rather than relying on heuristic or phenomenological arguments, we construct an exact framework in which wave dynamics and effective optical properties emerge directly from spacetime geometry (see also \cite{o1,o2,o3,o4,o5}). This approach allows waves propagation, reflection, and confinement to be interpreted as purely geometric effects.

\vspace{0.01cm}
\setlength{\parindent}{0pt}

This manuscript is structured as follows. In Sec.~\ref{sec:metric}, we establish the geometric framework by introducing a general static and spherically symmetric wormhole spacetime written in terms of a generalized radial coordinate that covers the entire manifold, and we provide a complete, coordinate-independent characterization of curvature through explicit evaluation of the Ricci and Kretschmann invariants. Building on this background, Sec.~\ref{sec:wave-propagation} analyzes the dynamics of a massless scalar field by reducing the covariant Klein-Gordon equation to a Helmholtz-like radial equation, thereby isolating the effects of redshift, curvature, and angular momentum into an effective potential. In Sec.~\ref{sec:ref-index}, this formulation is recast into an optical analogy through the introduction of a geometry-induced, frequency-dependent effective refractive index, emphasizing that the resulting refractive properties arise purely from spacetime curvature rather than any material optical response. The role of turning points is examined in Sec.~\ref{sec:turning-points}, where they can be identified as boundaries separating propagating and evanescent regions and are analyzed for different classes of lapse functions. Finally, Sec.~\ref{sec:conc} summarizes the results and discusses the broader implications of this unified geometric-optical framework for wave phenomena in Lorentz-violating wormhole spacetimes.

\section{Spacetime metric and curvature invariants}\label{sec:metric}

\setlength{\parindent}{0pt}

We consider a four-dimensional static and spherically symmetric spacetime described by the line element \cite{LV}:
\begin{equation}
ds^2
= -A(x)\,dt^2 + \frac{1}{A(x)}\,dx^2
+ r(x)^2\left(d\theta^2 + \sin^2\theta\,d\varphi^2\right),
\label{eq:metric}
\end{equation}
where $x\in(-\infty,\infty)$ is a generalized radial coordinate that covers the entire manifold of the wormhole, $A(x)$ denotes the lapse (redshift) function, and $r(x)$ is the areal radius. The latter is defined invariantly through the area $\mathcal{A}=4\pi r(x)^2$ of the spherical orbits of the rotation group and thus represents a genuine scalar determined by the spacetime geometry. Unlike Schwarzschild coordinates, in which the radial coordinate coincides with the areal radius, the parametrization \eqref{eq:metric} allows $r(x)$ to be an arbitrary smooth, nonvanishing function. This freedom is essential for the description of wormhole geometries, regular black holes, and other nonsingular compact objects. In particular, this parametrization avoids coordinate pathologies associated with the areal radius at the throat and allows both asymptotic regions of a wormhole to be described within a single chart. Throughout this section, a prime denotes differentiation with respect to $x$, and the metric signature $(-,+,+,+)$ is adopted. From Eq.~\eqref{eq:metric}, the nonvanishing components of the metric tensor and its inverse are
\begin{equation}
\begin{split}
&g_{tt}=-A(x), \quad g_{xx}=\frac{1}{A(x)}, \quad g_{\theta\theta}=r(x)^2, \quad g_{\varphi\varphi}=r(x)^2\sin^2\theta,\\
&g^{tt}=-\frac{1}{A(x)}, \quad g^{xx}=A(x), \quad g^{\theta\theta}=\frac{1}{r(x)^2}, \quad g^{\varphi\varphi}=\frac{1}{r(x)^2\sin^2\theta}.\label{metric-components}
\end{split}
\end{equation}
Since the metric functions depend solely on $x$, all nontrivial Levi-Civita connection coefficients involve derivatives with respect to this coordinate. Using the Levi-Civita connection definition \cite{book-1,book-2}
\(\Gamma^{\rho}{}_{\mu\nu}=\frac{1}{2}g^{\rho\sigma}\left(\partial_{\mu}g_{\nu\sigma}+\partial_{\nu}g_{\mu\sigma}-\partial_{\sigma}g_{\mu\nu}\right)\),
where Greek letters denote spacetime coordinates, one finds the independent nonvanishing Christoffel symbols:
\begin{align}
&\Gamma^{t}{}_{tx} = \frac{A'}{2A}, \quad \Gamma^{x}{}_{tt} = \frac{A A'}{2}, \quad \Gamma^{x}{}_{xx} = -\frac{A'}{2A}, \\
&\Gamma^{x}{}_{\theta\theta} = -A r r', \quad \Gamma^{x}{}_{\varphi\varphi} = -A r r' \sin^2\theta,\\
& \Gamma^{\theta}{}_{x\theta}= \Gamma^{\varphi}{}_{x\varphi}= \frac{r'}{r}, \quad \Gamma^{\varphi}{}_{\theta\varphi}= \cot\theta,\quad \Gamma^{\theta}{}_{\varphi\varphi}= -\cos\theta\,\sin\theta.
\end{align}
These expressions explicitly show that spacetime curvature originates from two independent geometric sources: variations of the lapse function $A(x)$, which encode gravitational time dilation and tidal effects in the $(t,x)$ sector, and variations of the areal radius $r(x)$, which control both the intrinsic curvature of the spherical sections and their embedding within the full spacetime. The Ricci tensor is defined by \cite{book-1,book-2}:
\begin{equation}
R_{\mu\nu}
=\partial_{\lambda}\Gamma^{\lambda}{}_{\mu\nu}
-\partial_{\nu}\Gamma^{\lambda}{}_{\mu\lambda}
+\Gamma^{\lambda}{}_{\mu\nu}\Gamma^{\sigma}{}_{\lambda\sigma}
-\Gamma^{\sigma}{}_{\mu\lambda}\Gamma^{\lambda}{}_{\nu\sigma}.
\end{equation}
Substitution of the above Christoffel symbols yields the nonvanishing components:
\begin{align}
R_{tt}&= \frac{1}{2}A A'' - A A'\frac{r'}{r}, \quad
R_{xx}= -\frac{A''}{2A}-\frac{2 r''}{r}-\frac{A'}{A}\frac{r'}{r}, \\
R_{\theta\theta}&= 1 - A r'^2 - A r r'' -  r r' A', \quad
R_{\varphi\varphi}= \sin^2\theta \, R_{\theta\theta}.
\end{align}
Contracting with the inverse metric, \(R = g^{\mu\nu} R_{\mu\nu}\), one finds, after cancellation of all terms quadratic in $A'(x)$ in the Ricci scalar,
\begin{equation}
R
= -\frac{1}{r^2}
\left[
2\bigl(A r'^2-1\bigr)
+ r^2 A''
+ 4 r \bigl(A r'\bigr)'
\right].
\label{eq:ricci_scalar}
\end{equation}
This structure reveals an important property: curvature divergences are governed by the behavior of the areal radius and by derivatives of the metric functions. A complete, coordinate-independent characterization of the curvature strength is provided by the Kretschmann scalar \(K = R_{\mu\nu\rho\sigma}R^{\mu\nu\rho\sigma}\) \cite{LV,book-2}, which captures the full magnitude of the Riemann tensor. The independent nonvanishing mixed-index components of the Riemann tensor are
\begin{align}
&R^{t}{}_{x t x} = -\frac{A''}{2A}, \quad
R^{t}{}_{\theta t \theta} = -\frac{1}{2} A' r r', \\
&R^{x}{}_{\theta x \theta} = -A r r''-\frac{A' r r'}{2}, \quad
R^{\theta}{}_{\varphi \theta \varphi}= \sin^2\theta \,(1 - A r'^2).
\end{align}
Using spherical symmetry and lowering indices with the metric, the Kretschmann scalar reduces to:
\begin{equation}
K
= 4 R_{t x t x}^2
+ 8 R_{t\theta t\theta}^2
+ 8 R_{x\theta x\theta}^2
+ 4 R_{\theta\varphi\theta\varphi}^2,
\end{equation}
which yields:
\begin{align}
K=\frac{r^4 \left(A''\right)^2
+2 r^2 \left(A' r'+2 A r''\right)^2
+2 r^2 \left(A'\right)^2 \left(r'\right)^2
+4 \left(A \left(r'\right)^2-1\right)^2}{r^4}.
\label{eq:kretschmann}
\end{align}
Each contribution admits a clear geometric interpretation. The first term primarily represents radial tidal curvature, arising from second derivatives of the lapse function and governing the relative acceleration of neighboring geodesics in the temporal-radial plane. The intermediate terms encode mixed tidal curvature effects generated by variations of the lapse function together with the radial embedding of the spherical sections. The final term encodes the intrinsic curvature of the two-spheres, depending on the deviation of the areal radius from its flat-space behavior through the combination \(1 - A(x)\,r'(x)^2\). For a wormhole geometry, the defining feature is the existence of a nonvanishing minimum of the areal radius at some $x=x_0$, satisfying:
\begin{equation}
r'(x_0)=0, \qquad r''(x_0)>0.
\end{equation}
These conditions ensure the presence of a smooth throat connecting two asymptotically distinct regions.

\section{Massless scalar wave propagation}\label{sec:wave-propagation}

\setlength{\parindent}{0pt}

The dynamics of a massless scalar field $\Phi$ propagating on the static and spherically symmetric background introduced in Eq.~\eqref{eq:metric} is governed by the covariant massless Klein-Gordon equation \cite{KG}:
\begin{equation}
\frac{1}{\sqrt{-g}} \, \partial_\mu \left( \sqrt{-g} \, g^{\mu\nu} \, \partial_\nu \Phi \right) = 0.
\label{OM1}
\end{equation}
This equation encodes the influence of spacetime curvature entirely through the metric tensor and its determinant. For the geometry under consideration, the square root of minus the metric determinant associated with Eq.~\eqref{eq:metric} takes the form $\sqrt{-g}=r(x)^2\sin\theta$. Exploiting the static character of spacetime and its spherical symmetry, the scalar field can be decomposed into temporal, radial, and angular parts using the standard separation ansatz \cite{KG}:
\begin{equation}
\Phi(t,x,\theta,\varphi)=e^{-i\omega t}\,Y_{\ell m}(\theta,\varphi)\,R(x),
\end{equation}
where $\omega$ denotes the wave frequency, while $Y_{\ell m}(\theta,\varphi)$ are the spherical harmonics on the unit two-sphere. These functions form a complete orthonormal basis and satisfy the eigenvalue equation \cite{KG}:
\begin{equation}
\Delta_{S^2}Y_{\ell m}=-\ell(\ell+1)Y_{\ell m},
\end{equation}
with $\ell=0,1,2,\ldots$ and $m=-\ell,\ldots,\ell$. Substituting $\Phi$ into the Klein-Gordon equation,  one finds that the angular and temporal dependencies decouple, leaving a single second-order differential equation for the radial function $R(x)$:
\begin{equation}
R''(x)
+ \left( \frac{A'(x)}{A(x)} + \frac{2 r'(x)}{r(x)} \right) R'(x)
+ \left( \frac{\omega^2}{A(x)^2}
- \frac{\ell(\ell+1)}{A(x) r(x)^2} \right) R(x) = 0,
\end{equation}
where a prime denotes differentiation with respect to the generalized radial coordinate $x$. The appearance of a first-derivative term reflects the nontrivial radial dependence of the background geometry and precludes a direct comparison with the flat-space wave equation. To cast the radial equation into a Helmholtz/Schr\"odinger-like form, which is particularly suited for physical interpretation, WKB analysis, and the development of optical analogies, we define
\begin{equation}
R(x)=\frac{\psi(x)}{r(x)\sqrt{A(x)}}.\label{field-redefinition}
\end{equation}
This redefinition allows for obtaining a second-order differential equation for $\psi(x)$ in the form:
\begin{equation}
\psi''(x)
+ \left[ \frac{\omega^2}{A(x)^2} - V_0(x) \right] \psi(x) = 0.
\label{eq:Schrodinger-form}
\end{equation}
Equation~\eqref{eq:Schrodinger-form} closely resembles an effective one-dimensional Schr\"odinger-like equation, with an energy-dependent effective kinetic factor $\omega^2/A(x)^2$ (reflecting the nontrivial redshift structure of the background), and an effective potential $V_0(x)$ that encapsulates the combined effects of curvature, redshift, and angular momentum. The explicit form of which is given by:
\begin{equation}
V_0(x)= \frac{r''(x)}{r(x)}
- \frac{A'(x)^2}{4A(x)^2}
+ \frac{A''(x)}{2A(x)}
+ \frac{r'(x)A'(x)}{r(x)A(x)}
+ \frac{\ell(\ell+1)}{A(x)\,r(x)^2}.
\label{eq:V0-canonical}
\end{equation}
Each term in $V_0(x)$ admits a clear geometric interpretation: the derivatives of $r(x)$ encode variations of the areal radius, the derivatives of $A(x)$ reflect gravitational redshift effects, and the final term represents the centrifugal barrier associated with angular momentum. In this form, the scalar wave problem reduces to the study of a one-dimensional scattering problem in an effective potential, providing a powerful framework for analyzing wave propagation, turning points, resonances, and optical properties of the underlying spacetime geometry. This formulation will serve as the basis for the geometric-optical interpretation developed in the following sections.

\section{Effective refractive index}
\label{sec:ref-index}

\setlength{\parindent}{0pt}

The Schr\"odinger-like radial equation~(\ref{eq:Schrodinger-form}) can be naturally recast into its Helmholtz optical analogue, providing a rigorous and physically transparent framework for describing scalar-wave propagation in curved spacetime. Within this framework, the radial dynamics can be interpreted as those of a monochromatic wave propagating through an effective medium whose properties are determined entirely by the background geometry. This analogy is not merely heuristic; rather, it follows directly from the structure of the covariant wave equation in a static, spherically symmetric spacetime. For such backgrounds, the radial equation can be written in a Helmholtz-like form by introducing a position- and frequency-dependent effective wavenumber $k_{\mathrm{eff}}(x,\omega)$~\cite{o1,o2,o3,o4,o5}, thereby obtaining:
\begin{equation}
\psi''(x) + k_{\mathrm{eff}}(x,\omega)^2\,\psi(x) = 0 ,
\label{4.1}
\end{equation}
where $k_{\mathrm{eff}}$ incorporates all geometric and angular-momentum contributions. Its explicit expression is
\begin{equation}
\begin{split}
k_{\mathrm{eff}}(x,\omega)^2= \frac{\omega^2}{A(x)^2}
-\quad&\frac{r''(x)}{r(x)}
+\frac{A'(x)^2}{4A(x)^2}
-\frac{A''(x)}{2A(x)}\\-\frac{r'(x)A'(x)}{r(x)A(x)}
&-\frac{\ell(\ell+1)}{r(x)^2A(x)} .
\end{split}
\label{4.2}
\end{equation}
The leading term represents the gravitational redshift frequency, whereas the remaining contributions define an effective potential arising from the curvature of spacetime and the centrifugal barrier associated with angular momentum $\ell$. Lorentz-violating effects enter implicitly through the functional form of the lapse function \(A(x)\), which encodes deviations from standard general relativistic solutions. It is then convenient to rewrite Eq.~\eqref{4.1} in the form of a generalized Helmholtz-type equation \cite{o1,o2,o3,o4,o5},
\begin{equation}
\psi''(x) + \omega^2\,n(\omega,x)^2\,\psi(x) = 0,
\label{4.3}
\end{equation}
which allows one to identify an effective refractive index $n(\omega,x)$ induced by spacetime geometry. Direct comparison yields:
\begin{align}
n(\omega,x)^2
&= \frac{1}{A(x)^2}
+\frac{1}{\omega^2}\Bigg[
\frac{A'(x)^2}{4A(x)^2}
-\frac{r''(x)}{r(x)}
-\frac{A''(x)}{2A(x)} \nonumber\\
&\qquad -\frac{r'(x)A'(x)}{r(x)A(x)}
-\frac{\ell(\ell+1)}{r(x)^2A(x)}
\Bigg] ,
\label{ref-index}
\end{align}
This expression shows that the effective refractive index \(n(\omega,x)^2\) depends explicitly on both position and frequency and may develop singularities associated with the lapse function \(A(x)=0\) and/or the areal radius \(r(x)=0\). In the optical analogy, such divergences signal a close connection between Killing horizons and the refractive-index structure. In particular, the divergence of \(n(\omega,x)\) at \(A(x)=0\) reflects the breakdown of the static optical description at horizons, consistent with the causal trapping of null rays in general relativity. The refractive-index singularities do not, in general, correspond to genuine curvature singularities; rather, they indicate the breakdown of the geometric-optics approximation and are associated either with Killing horizons (where \(n(\omega,x)\) diverges) or with classical turning points in wave propagation (where \(n(\omega,x)^2=0\)). The presence of \(\omega^{-2}\) terms encodes dispersion-like effects induced purely by spacetime curvature and Lorentz-violating geometry, which become negligible in the high-frequency limit but play a crucial role at finite wavelengths. The spatial variation of \(n(\omega,x)\) therefore governs the local propagation properties of the scalar field, including reflection, transmission, and tunneling phenomena. Within this optical picture, regions where \(n(\omega,x)^2 > 0\) support oscillatory (propagating) solutions, whereas regions where \(n(\omega,x)^2 < 0\) correspond to evanescent behavior \cite{o3,o5}. The transition between these regimes is determined by the turning-point condition (i.e. \(n(\omega,x)^2 = 0\)), which plays the role of an effective optical boundary that separates the classically allowed and forbidden regions for wave propagation. From this perspective, the curvature-induced potential acts as an optical barrier that controls scattering, partial transmission, and wave confinement. 

\section{Turning points and optical interpretation}
\label{sec:turning-points}

\setlength{\parindent}{0pt}

Within the optical framework developed in Sec.~\ref{sec:ref-index}, turning points acquire a clear geometric and physical interpretation. They mark the boundaries between regions where scalar waves propagate oscillatory and regions where they become evanescent, thereby governing reflection, transmission, and spatial confinement. In curved spacetimes, these turning points are not fixed a priori but are determined by the combined effects of the wave frequency, angular momentum, spacetime curvature, and Lorentz-violating contributions encoded in the lapse function. Starting from the Helmholtz-like equation~\eqref{4.3}, with the geometry-induced refractive index $n(\omega,x)$, the effective wavenumber is given (with $c=1$) by $k_{\rm eff}(\omega,x)^2=\omega^2 n(\omega,x)^2$. The turning points $x_t$ are defined by the condition $n(\omega,x_t)^2=0$. At these points, the WKB approximation breaks down as the solution transitions from oscillatory to exponential behavior. In optical terms, the turning points act as effective interfaces separating transparent and opaque regions of the curved spacetime, analogous to reflection surfaces in inhomogeneous optical media.

\subsection{Constant lapse function: no Killing horizons}

We first consider geometries with a constant lapse function \cite{LV}, \(A(x)=a_1,\), which we set to unity (\(a_1=1\)) without loss of generality. This case describes a static, horizonless spacetime symmetric about the throat, and corresponds to a two-sided wormhole geometry. The areal radius is
\begin{equation}
r(x)=\sqrt{a^2+\frac{x^2}{1-\eta}}\,; \qquad \eta<1,\label{CAR}
\end{equation}
where $a$ is the throat radius and $\eta$ quantifies Lorentz-violating parameter/deviations. Since the lapse function never vanishes and the areal radius remains strictly positive, no singularities arise in the refractive index (\ref{ref-index}); consequently, no Killing horizons are present in this geometry.
\begin{figure*}[!htbp]
    \centering
    \includegraphics[width=0.28\linewidth]{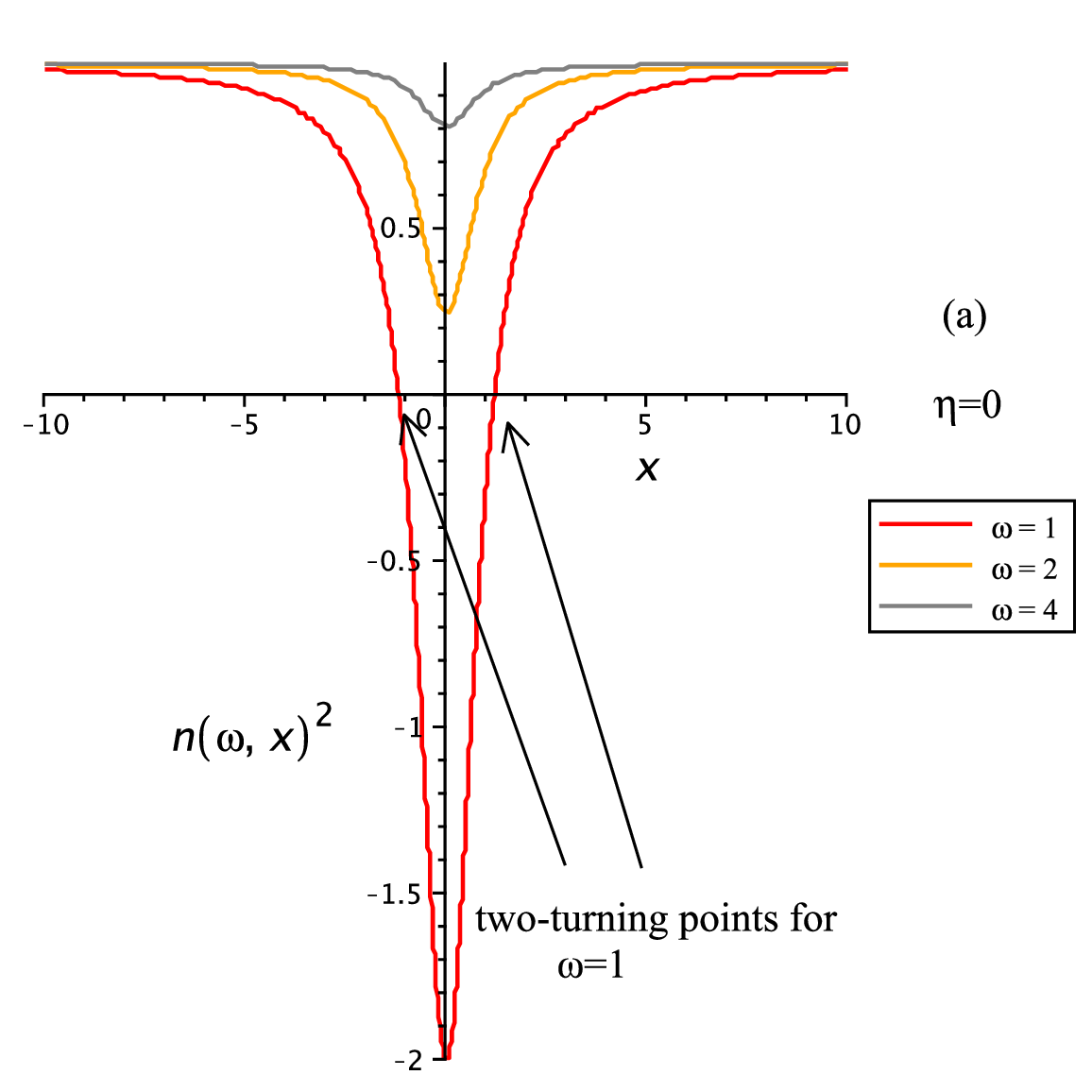}
    \includegraphics[width=0.28\linewidth]{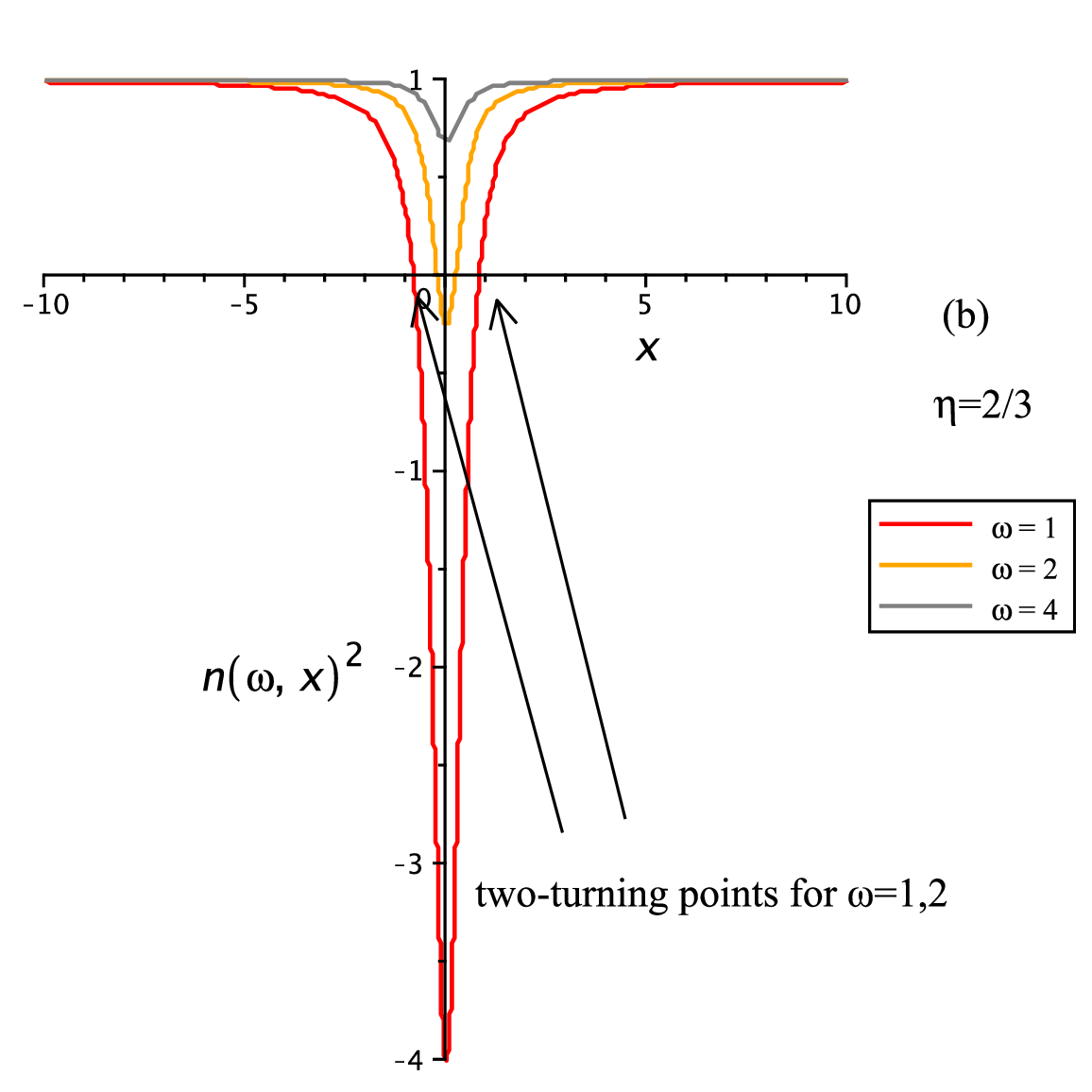}
    \includegraphics[width=0.28\linewidth]{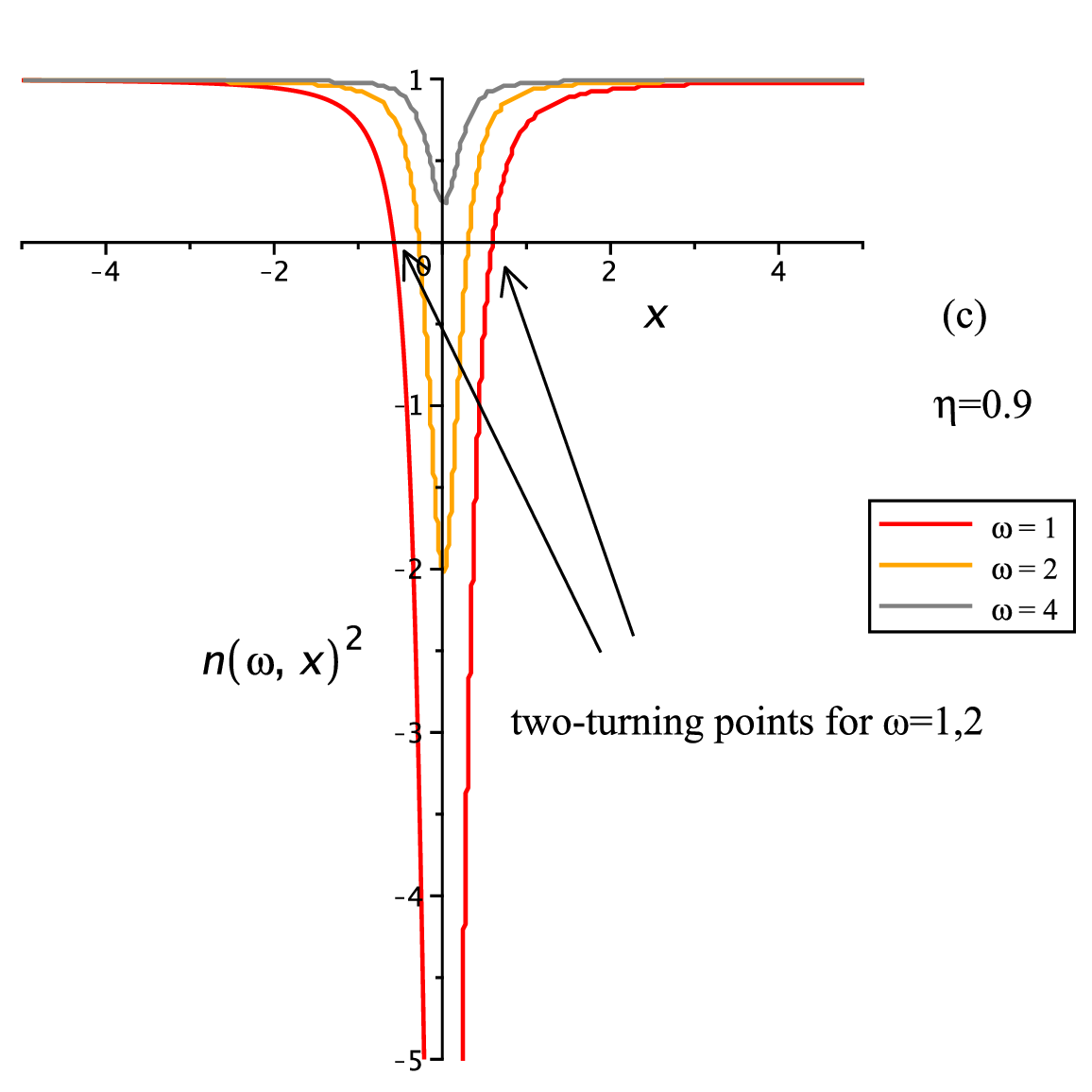}
    \caption{\fontsize{7.6}{8.6}\selectfont 
    Plots of \(n(\omega,x)^2\) versus $x$ for $\ell=a=1$ and different Lorentz-violating parameters \(\eta=0,2/3,0.9\) at \(\omega=1,2,4\).\label{fig:1} 
    Table~\ref{tab:turning_points1} reports the corresponding turning point locations.}
\end{figure*}

\begin{table}[!htbp]
\centering
{\fontsize{8}{9}\selectfont 

\caption{Turning points \(x_t\) defined by \(n(\omega,x)=0\), and Killing horizon locations \(x_h\) defined by \(A(x)=0\), for the constant lapse geometry with $\ell=a=1$ and \(\eta=0, 2/3, 0.9\) at \(\omega=1,2,4\).}
\label{tab:turning_points1}
\vspace{0.2cm}

\begin{tabular}{cccc}
\toprule
$\eta$ & $\omega$ & $x_t$ & $x_h$ \\ 
\midrule

0.0   & 1 & $\displaystyle \overbrace{-1.2,\,1.2}^{\text{appear in Fig.~\ref{fig:1}}}$ & none \\
      & 2 & none & none \\
      & 4 & none & none \\[0.2cm]

2/3   & 1 & $\displaystyle \overbrace{-0.8,\,0.8}^{\text{appear in Fig.~\ref{fig:1}}}$ & none \\
      & 2 & $\displaystyle \overbrace{-0.2,\,0.2}^{\text{appear in Fig.~\ref{fig:1}}}$ & none \\
      & 4 & none & none \\[0.2cm]

0.9   & 1 & $\displaystyle \overbrace{-0.6,\,0.6}^{\text{appear in Fig.~\ref{fig:1}}}$ & none \\
      & 2 & $\displaystyle \overbrace{-0.3,\,0.3}^{\text{appear in Fig.~\ref{fig:1}}}$ & none \\
      & 4 & none & none \\

\bottomrule
\end{tabular}
} 
\end{table}

Substitution into the general expression for the effective wavenumber yields
\begin{equation}
k_{\rm eff}(\omega,x)^2=
\omega^2
-\frac{\ell(\ell+1)(1-\eta)}{a^2(1-\eta)+x^2}
-\frac{a^2(1-\eta)}{\left[a^2(1-\eta)+x^2\right]^2}.
\label{eq:keff_constant_extended}
\end{equation}
The effective wavenumber is finite and symmetric under $x\to -x$, reflecting the regularity of the geometry. In Fig.~\ref{fig:1}, we plot \(n(\omega,x)^2=k_{\rm eff}(\omega,x)^2/\omega^2\) for different Lorentz-violating parameters \(\eta=0,2/3,9/10\) and frequencies \(\omega=1,2,4\). Far from the throat, \(|x|\to\infty\), the refractive index approaches unity, and the spacetime becomes optically transparent. Turning points occur at locations where \(n(\omega,x)=0\), marking the transition between oscillatory and evanescent wave propagation. Low-frequency modes encounter turning points and are reflected at finite distances from the throat, whereas high-frequency modes penetrate deeply into the geometry and traverse the wormhole with weak reflection. A comparison between Figs.~\ref{fig:1}(a)-(c) shows that increasing the Lorentz-violating parameter enhances the effective angular momentum barrier, shifting the turning points and narrowing the transmission window. Therefore, Lorentz violation modifies wave propagation without introducing horizons or instabilities.

\subsection{Linear lapse function: one Killing horizon}

We now consider a geometry characterized by a linearly varying lapse function \cite{LV},
\begin{equation}
A(x) = 1 + \chi x,\label{LLF}
\end{equation}
where $\chi$ defines an effective acceleration scale along the $x$-direction. The linear dependence of $A(x)$ introduces a Lorentz-violating gradient in time dilation, breaking the usual $x \to -x$ symmetry and leading to direction-dependent propagation effects for waves in this background. The corresponding areal radius $r(x),$ which measures the proper circumferential distance, takes the general form:
\begin{equation}
r(x) = \sqrt{a^2 + \frac{2(2N_1 - 1)\left[-1 - N_1 \chi x + (1 + \chi x)^{N_1}\right]}{(N_1 - 1) N_1 \chi^2}},\label{LAR1}
\end{equation}
with $N_1 = (\eta - 2)/(2\eta - 2)$ determined by the polytropic index $\eta$ of the effective matter content. The constant $a^2$ ensures regularity at $x = 0$, preventing a singularity in the areal radius. Magalhães et al.~\cite{LV} reported that \(r(x)^2\) is symmetric with respect to the throat only for \(N_1=2\to\eta=2/3\); any other value of \(\eta\) renders the areal radius asymmetric and potentially complex when \((1+\chi x)<0\), which typically occurs in the causally disconnected region beyond the Killing horizon. Consequently, adopting \(N_1=2\) in (\ref{LAR1}) yields the symmetric areal radius:
\begin{equation}
r(x) = \sqrt{a^2 + 3x^2}.\label{LAR}
\end{equation}
The linear lapse function (\ref{LLF}) introduces a single Killing horizon at \(x_h=-1/\chi\), where \(A(x_h)=0\). No additional horizons arise, since the areal radius remains strictly positive, \(r(x)>0\), and thus does not generate further singularities in the refractive index \(n(\omega,x)\).

\vspace{0.01cm}
\setlength{\parindent}{0pt}

\begin{figure*}[ht!]
    \centering
    \includegraphics[width=0.28\linewidth]{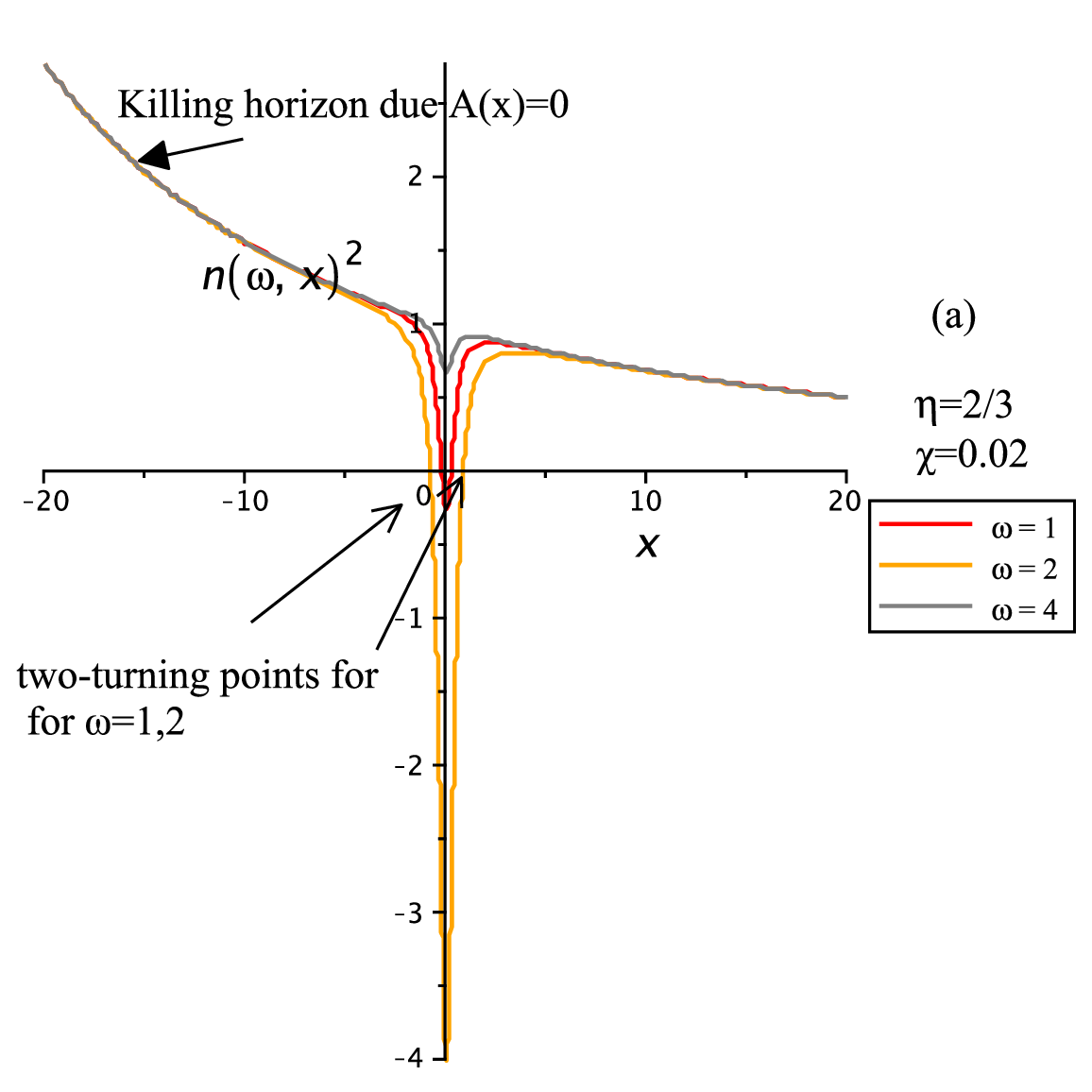}
    \includegraphics[width=0.28\linewidth]{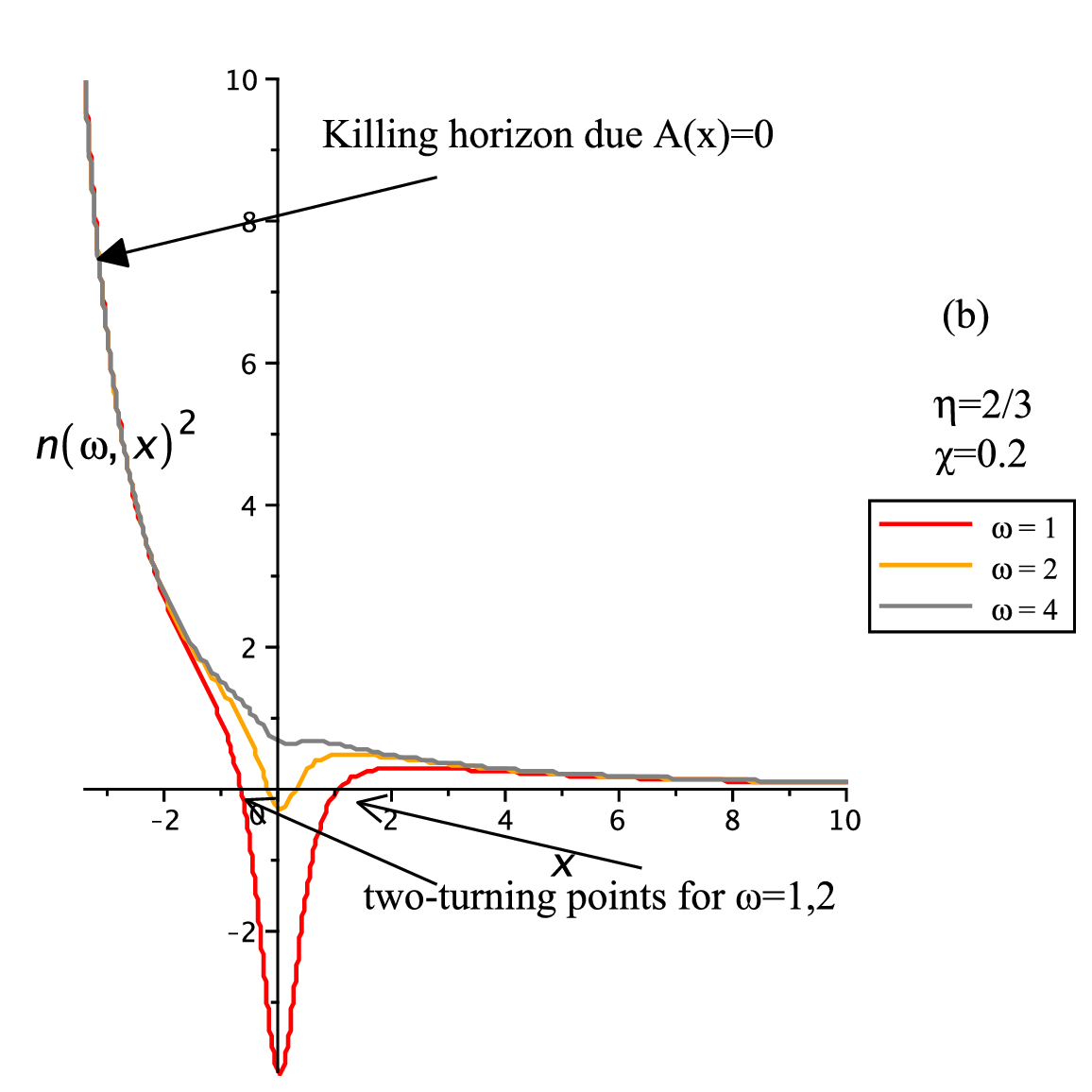}
    \includegraphics[width=0.28\linewidth]{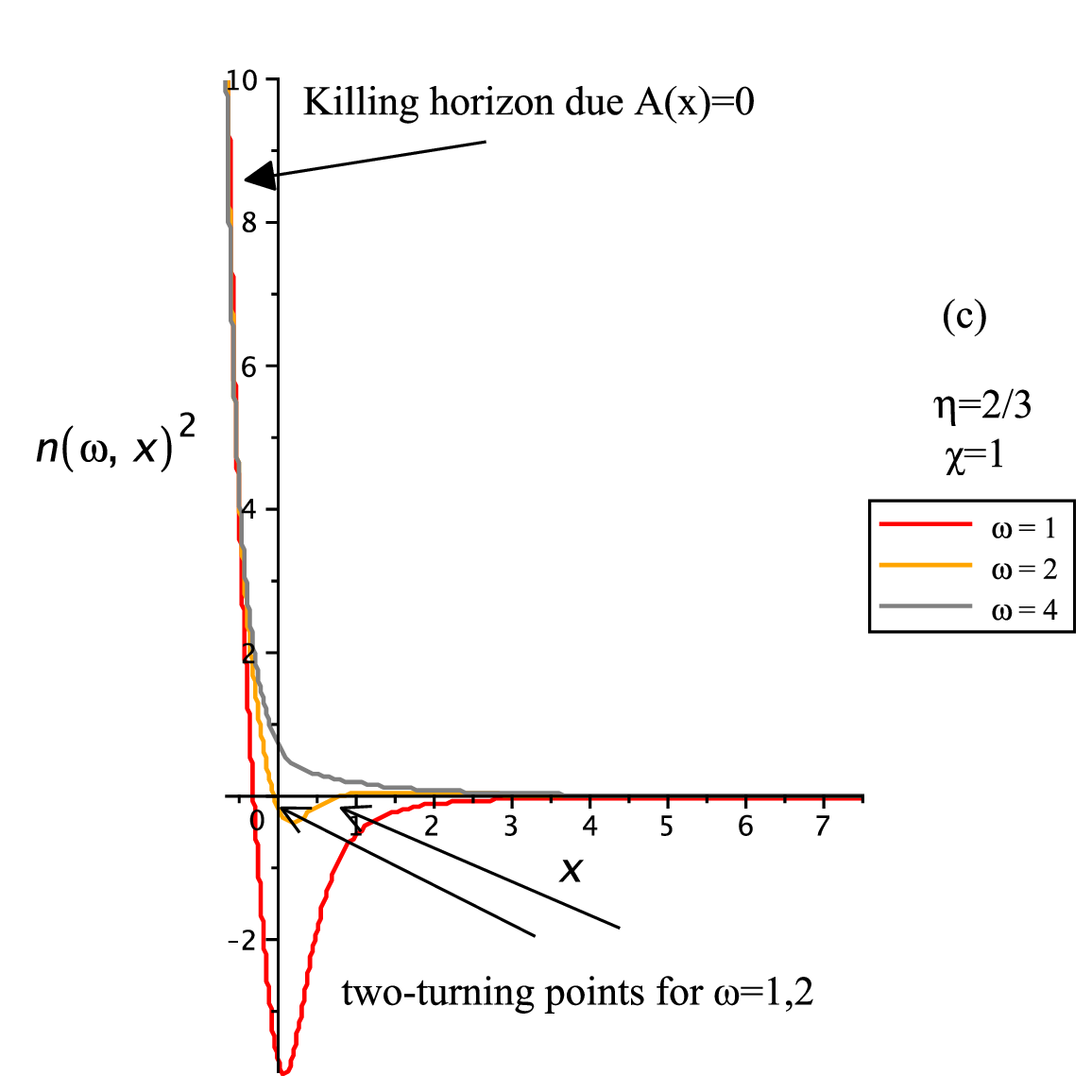}
    \caption{\fontsize{7.6}{8.6}\selectfont Plots of \(n(\omega,x)^2\) versus $x$ for $\ell = a = 1$, \(\eta=2/3\), and \(\omega=1,2,4\), for different Rindler-type acceleration values \(\chi=0.02,0.2,1\). \label{fig:2}
    Table~\ref{tab:turning_points2} reports the corresponding turning points and Killing horizon locations.}
\end{figure*}

In Fig.~\ref{fig:2}, we plot the refractive index squared (\ref{ref-index}) as a function of \(x\) for $\ell = a = 1$, \(\omega=1,2,4\), and the Lorentz-violating parameter \(\eta=2/3\), for different values of the Rindler-type acceleration \(\chi=0.02,0.2,1\). The figure shows that increasing \(\chi\) shifts the Killing horizon toward the throat from the negative-$x$ side, indicating enhanced reflection of waves approaching from the left, while waves propagating toward positive $x$ encounter an increasingly transparent region and can traverse the wormhole with minimal reflection. The asymmetry of the lapse function in (\ref{LLF}) thus induces a direction-dependent refractive profile: waves propagating toward increasing $x$ experience an effective blueshift, which enhances transmission, whereas waves incident from decreasing $x$ encounter a refractive barrier that suppresses propagation. This behavior closely parallels graded-index optical media, where spatial variations of the refractive index produce direction-dependent scattering. In the present case, the Lorentz-violating linear lapse generates an analogous phenomenon, demonstrating that even a simple linear lapse profile can lead to rich asymmetric wave dynamics in curved spacetimes.

\begin{table}[!htbp]
\centering
{\fontsize{8}{9}\selectfont 

\caption{Turning points $x_{t}$ and Killing horizons $x_{h}$ (boxed) defined by \(A(x)=0\), for the linear lapse geometry, with $\ell = a = 1$ and \(\eta=2/3\) for different values of \(\chi\) and \(\omega\).}
\label{tab:turning_points2}
\vspace{0.2cm}

\begin{tabular}{ccccc}
\toprule
$\chi$ &\quad $\omega$ &\quad $x_{t}$ & \quad$x_{h}$  \\ 
\midrule

0.02 & 1 & $\displaystyle \overbrace{-0.8,\, 0.8}^{\text{appear in Fig.~\ref{fig:2}}}$ & $\fbox{-50}$ \\ 
     & 2 & $\displaystyle \overbrace{-0.2,\, 0.2}^{\text{appear in Fig.~\ref{fig:2}}}$ & $=$ \\ 
     & 4 & none & $=$ \\[0.2cm]

0.2  & 1 & $\displaystyle \overbrace{-0.7,\,1.1}^{\text{appear in Fig.~\ref{fig:2}}}$ & $\fbox{-5}$ \\  
     & 2 & $\displaystyle \overbrace{-0.2,\,0.3}^{\text{appear in Fig.~\ref{fig:2}}}$ & $=$ \\ 
     & 4 & none & $=$ \\[0.2cm]

1    & 1 & $\displaystyle \overbrace{-0.3,\,7.1}^{\text{appear in Fig.~\ref{fig:2}}}$ & $\fbox{-1}$ \\ 
     & 2 & $\displaystyle \overbrace{-0.06,\,0.6}^{\text{appear in Fig.~\ref{fig:2}}}$ & $=$ \\ 
     & 4 & none & $=$ \\ 

\bottomrule
\end{tabular}

}
\end{table}

For the linear-lapse wormhole geometry defined by \eqref{LLF} and the areal radius \eqref{LAR}, corresponding to \(\eta=2/3\), the admissibility of the parameter \(\chi\) is dictated not by geometric regularity but by wave-optical and causal considerations relevant to scalar propagation. Since \(r(x)^{2}=a^{2}+3x^{2}>0;\,\forall x,\) spatial geometry is everywhere regular and imposes no restriction on \(\chi\). Moreover, because \(A'(x)=\chi=\mathrm{const}\) and \(A''(x)=0\), curvature invariants remain finite for finite \(x\), and the vanishing of the lapse does not correspond to a curvature singularity. Instead, the surface \(A(x_{h})=0\) at \(x_{h}=-1/\chi\) represents a Killing horizon. The relevant constraint on \(\chi\) arises from the effective refractive index \(n(\omega,x)^{2}\), whose dominant geometric-optics contribution scales as \(A(x)^{-2}\). At the horizon, the refractive index diverges, \(n(\omega,x)^{2}\to\infty\), signaling a causal boundary beyond which the geometric-optics description ceases to be physically meaningful. To ensure that scalar waves are reflected before reaching this boundary, the classical turning point \(x_{t}\), defined by \(n(\omega,x_{t})^{2}=0\), must satisfy \(x_{t}>x_{h}\). In the regime \(|\chi x_{t}|\ll 1\), and using the asymptotic expression for the effective radial wavenumber given in Eq.~\eqref{4.2}, the turning point on the negative-$x$ branch is located at
\[
|x_{t}|\simeq \frac{\sqrt{C_{\ell}}}{\omega},
\]
where \(C_{\ell}=3/4+\ell(\ell+1)/3\). Imposing the condition \(x_{t}>-1/\chi\) yields the upper bound
\begin{equation}
\chi<
\frac{\omega}
{\sqrt{\,3/4+\ell(\ell+1)/3\,}},
\end{equation}
which constitutes a constraint on the linear lapse gradient. Exceeding this bound does not introduce a geometric pathology; rather, it shifts the horizon beyond the turning point, removing the evanescent barrier and suppressing wave reflection. The admissible values of \(\chi\) are therefore controlled by the combined effects of Lorentz-violating redshift, angular-momentum-induced centrifugal terms, and the frequency-dependent nature of wave propagation, rather than by spacetime regularity alone.

\subsection{Quadratic lapse function: two Killing horizons.}

\begin{figure*}[!htbp]
    \centering
    \includegraphics[width=0.28\linewidth]{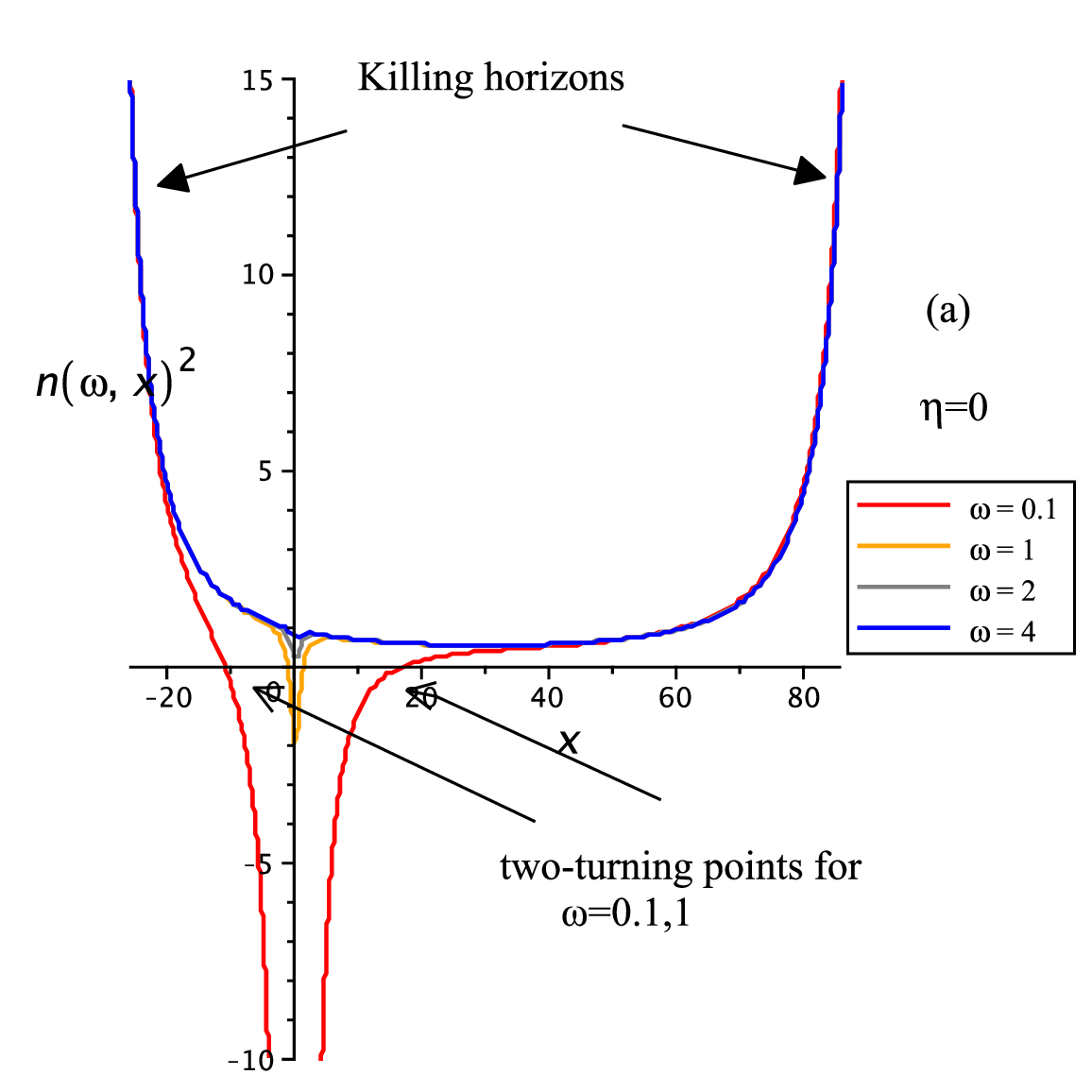}
    \includegraphics[width=0.28\linewidth]{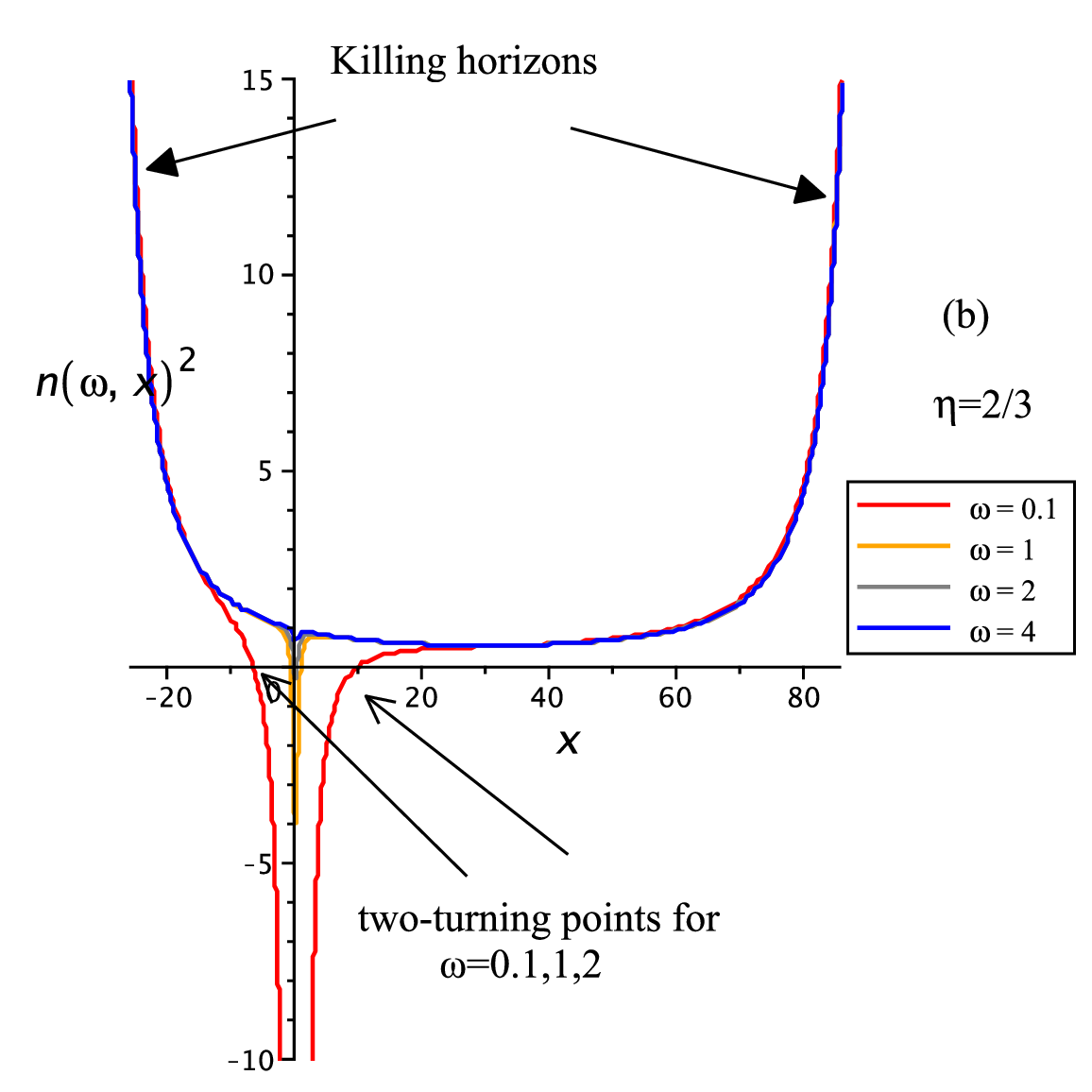}
    \includegraphics[width=0.28\linewidth]{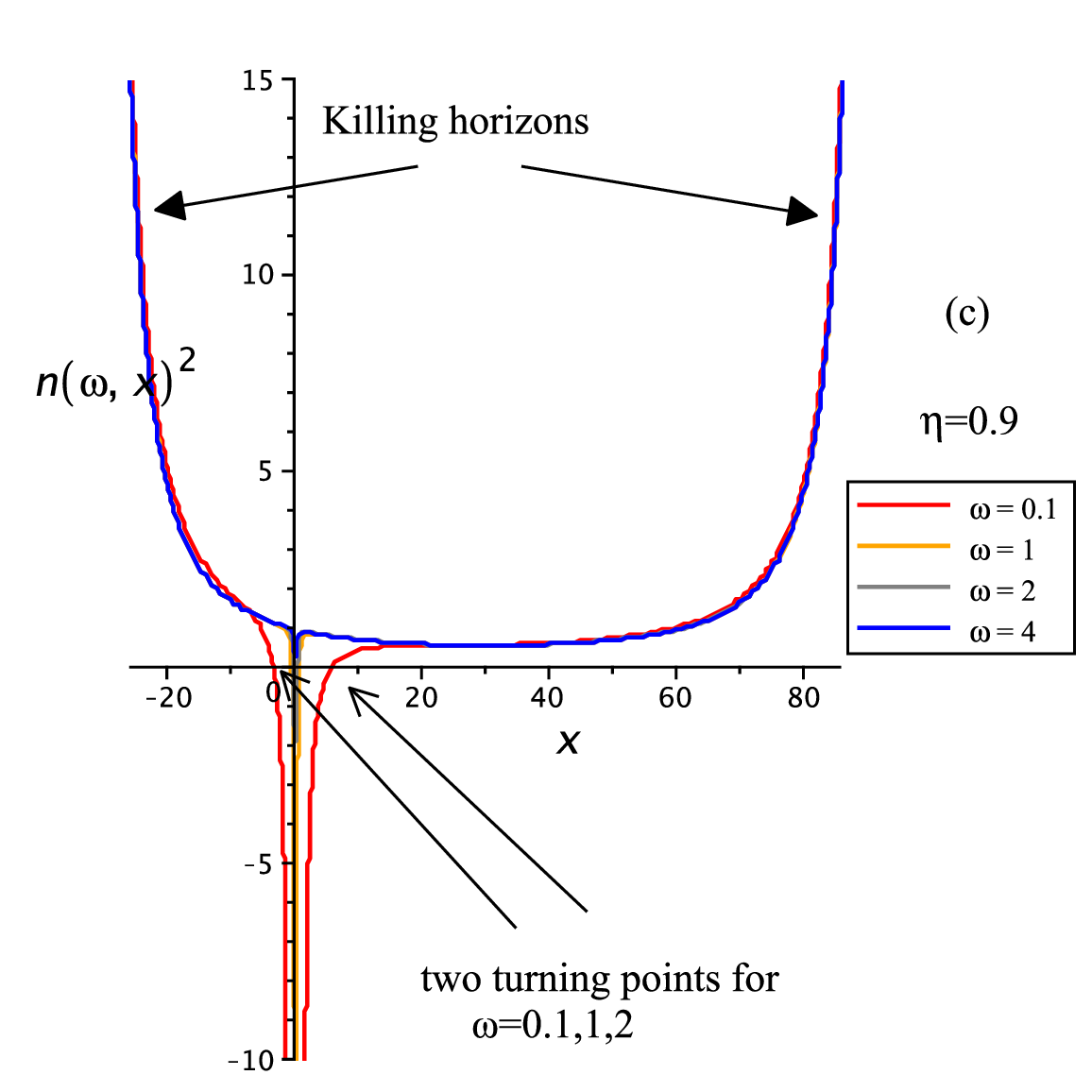}
    \caption{\fontsize{7.6}{8.6}\selectfont 
    Effective refractive index squared \(n(\omega,x)^2\) for the quadratic-lapse geometry, with $\ell = a = 1$. When \(n(\omega,x)^2<0\) within the plotted ranges, indicating evanescent modes. Table~\ref{tab:turning_points3} summarizes the turning points (a sample is shown on each panel) and Killing horizons for \(\omega=0.1,1,2,4\).}
    \label{fig:3}
\end{figure*}

\begin{table}[!htbp]
\centering
{\fontsize{8}{9}\selectfont 

\caption{For the quadratic lapse function, the killing horizons \(x_h\) (boxed) indicate the physically admissible causal domain. Over braced turning points \(x_t\) labeled as ``appear in Fig.~\ref{fig:3}'' lie within the admissible region, whereas those labeled ``not within Fig.~\ref{fig:3} range'' are located beyond the causal horizons and are physically inadmissible.}
\label{tab:turning_points3}
\vspace{0.2cm}

\begin{tabular}{ccccc}
\toprule
$\eta$ &\quad $\omega$ & \quad$x_{t}$ &\quad $x_{h}$  \\ 
\midrule

0   & 0.1 & $-188.1,\displaystyle \overbrace{-11.0,\,15.9}^{\text{appear in Fig.~\ref{fig:3}}},\,273.2$ & $\fbox{-32.45,\,92.45}$ \\ 
    & 1 & $-2078.3,\displaystyle \overbrace{-1.2,\,1.2}^{\text{appear in Fig.~\ref{fig:3}}},2168.2$ & $=$ \\ 
    & 2 & $\displaystyle \overbrace{-4198.6,\,4288.6}^{\text{not within Fig.~\ref{fig:3} range}}$ & $=$ \\ 
    & 4 & $\displaystyle \overbrace{-8440.8,\,8530.8}^{\text{not within Fig.~\ref{fig:3} range}}$ & $=$ \\[0.2cm]

2/3 & 0.1 & $-182.8,\displaystyle \overbrace{-6.5,\,9.4}^{\text{appear in Fig.~\ref{fig:3}}},\,269.9$ & $=$ \\
    & 1 & $-2077.8,\displaystyle \overbrace{-0.8,\,0.8}^{\text{appear in Fig.~\ref{fig:3}}},\,2167.8$ & $=$ \\
    & 2 & $-4198.4,\displaystyle \overbrace{-0.2,\,0.2}^{\text{appear in Fig.~\ref{fig:3}}},\,4288.4$ & $=$ \\ 
    & 4 & $\displaystyle \overbrace{-8440.6,\,8530.6}^{\text{not within Fig.~\ref{fig:3} range}}$ & $=$ \\[0.2cm]

0.9 & 0.1 & $-180.9,\displaystyle \overbrace{-3.4,\,5.6}^{\text{appear in Fig.~\ref{fig:3}}},268.7$ & $=$ \\ 
    & 1 & $-2077.6,\displaystyle \overbrace{-0.6,\,0.6}^{\text{appear in Fig.~\ref{fig:3}}},2167.6$ & $=$ \\ 
    & 2 & $-4198.3,\displaystyle \overbrace{-0.3,\,0.3}^{\text{appear in Fig.~\ref{fig:3}}},4288.3$ & $=$ \\ 
    & 4 & $\displaystyle \overbrace{-8440.6,\,8530.6}^{\text{not within Fig.~\ref{fig:3} range}}$ & $=$ \\ 

\bottomrule
\end{tabular}

} 
\end{table}

Finally, we consider a quadratic lapse function \cite{LV},
\begin{equation}
A(x)=1+\chi x-\frac{\Lambda_e}{3}x^2,
\label{QLF}
\end{equation}
associated with an effective cosmological parameter $\Lambda_e \ll 1$. The corresponding areal radius is
\begin{equation}
r(x)=\sqrt{a^2+\tilde{\eta}\,x^2}, \qquad 
\tilde{\eta}=\frac{6-a^2\Lambda_e(7\eta-2)}{6(1-\eta)}>0.
\label{QAR}
\end{equation}
At this point, one should note that $\tilde{\eta}>0$ since $\eta<1$ and $\Lambda_e \ll 1$. The quadratic lapse function suggests a class of solutions endowed with an effective cosmological constant $\Lambda \sim 10^{-52}\,\mathrm{m}^{-2} \sim 10^{-123}\,\ell_p^{-2}$, similar to de Sitter and anti–de Sitter asymptotics \cite{LV}. Consequently, the areal radius introduces no real singularities in the refractive index, since
\[
r(x)=0 \;\Rightarrow\; x=\pm i a \sqrt{\frac{6(1-\eta)}{6-\Lambda_e a^2(7\eta-2)}} \in \mathbb{C},
\]
which implies that no additional horizons arise from this condition. The quadratic lapse function \eqref{QLF} instead introduces two Killing horizons located at
\[
x_h=\frac{3\chi \pm \sqrt{9\chi^2+12\Lambda_e}}{2\Lambda_e}.
\]
Determining the physically admissible propagation region therefore requires a careful assessment of causal accessibility and the locations of wave-optical turning points. Figure~\ref{fig:3}, together with Table~\ref{tab:turning_points3}, shows (for $\ell = a = 1$ and $\eta = 0,\,2/3,\,0.9$) two cosmological Killing horizons located at $x_h=-32.45$ and $x_h=92.45$, defining the causally accessible propagation region. The turning points exhibit the following behavior:
\begin{enumerate}
\item For $\eta=0$, two turning points at $x_t=-11.0,\,15.9$ and $x_t=-1.2,\,1.2$ (labeled as ``appear Fig.~\ref{fig:3}'' in Table~\ref{tab:turning_points3}) for $\omega=0.1$ and $\omega=1$, respectively, lie within the two Killing horizons and are therefore physically admissible. In contrast, for $\omega=2$ and $\omega=4$, the corresponding turning points lie far outside the cosmological horizons and are thus physically inadmissible, as shown in Fig.~\ref{fig:3}(a).
    
\item For $\eta=2/3$, two turning points at $x_t=-6.5,\,9.4$, $x_t=-0.8,\,0.8$, and $x_t=-0.2,\,0.2$ (labeled as ``appear Fig.~\ref{fig:3}'' in Table~\ref{tab:turning_points3}) for $\omega=0.1$, $\omega=1$, and $\omega=2$, respectively, lie within the two Killing horizons and are therefore physically admissible. In contrast, for $\omega=4$, the turning points lie far outside the cosmological Killing horizons and are thus physically inadmissible, as shown in Fig.~\ref{fig:3}(b).
    
\item For $\eta=0.9$, two turning points at $x_t=-3.4,\,5.6$, $x_t=-0.6,\,0.6$, and $x_t=-0.3,\,0.3$ (labeled as ``appear Fig.~\ref{fig:3}'' in Table~\ref{tab:turning_points3}) for $\omega=0.1$, $\omega=1$, and $\omega=2$, respectively, lie within the two Killing horizons and are therefore physically admissible. However, for $\omega=4$, the turning points lie far outside the cosmological Killing horizons and are thus physically inadmissible, as shown in Fig.~\ref{fig:3}(c).
\end{enumerate}
From the optical viewpoint, this spacetime behaves as a confining medium supporting trapped or quasi-bound modes. The confinement arises from the global geometric structure rather than fine-tuning of the wave frequency. Turning points therefore provide a unified geometric criterion for wave propagation, reflection, and confinement in Lorentz-violating curved spacetimes. Interpreted through the effective refractive index, they encode how curvature and Lorentz violation reshape the optical properties of the vacuum, linking frequency-dependent scattering, directional asymmetry, and geometric confinement within a single coherent framework.

\section{Summary and Discussion}
\label{sec:conc}

\setlength{\parindent}{0pt}

In this work, we developed a geometric and wave-optical framework for the propagation of massless scalar waves in static, spherically symmetric Lorentz-violating wormhole spacetimes. Starting from a general metric parametrized by an arbitrary lapse function and areal radius, we derived curvature invariants and established regularity conditions at the wormhole throat. By solving the covariant Klein-Gordon equation, we cast the radial dynamics into a Helmholtz-like form with an effective curvature-induced potential. This formulation naturally led to a position- and frequency-dependent effective refractive index, allowing the curved vacuum to be interpreted as an inhomogeneous optical medium whose properties are determined by the underlying geometry and Lorentz-violating structure.

\vspace{0.01cm}

A central result of our analysis is the distinction between geometric (curvature) singularities and optical singularities. While curvature invariants remain finite throughout the spacetime, divergences of the effective refractive index correspond to Killing horizons and mark boundaries of the geometric-optics regime rather than true curvature pathologies. In contrast, turning points defined by the vanishing of the refractive index represent genuine wave-optical interfaces separating oscillatory and evanescent regions. These turning points govern reflection, transmission, and confinement of scalar waves and depend sensitively on the wave frequency, angular momentum, and Lorentz-violating parameters encoded in the lapse function.

\vspace{0.01cm}

We applied this framework to constant, linear, and quadratic lapse profiles, revealing a rich variety of wave-optical phenomena. Constant-lapse geometries are horizonless and symmetric, supporting frequency-dependent reflection from curvature-induced barriers and near-transparent transmission at high frequencies. Linear-lapse profiles break spatial symmetry and introduce direction-dependent refractive-index gradients, leading to asymmetric scattering analogous to graded-index optical media. Quadratic-lapse profiles can generate multiple Killing horizons and confinement regions, where global geometric structure produces trapped or quasi-bound modes without extreme fine-tuning of the wave frequency. Importantly, admissibility constraints on Lorentz-violating parameters were shown to arise from wave-optical causality conditions, rather than from curvature regularity alone.

\vspace{0.01cm}

Our results demonstrate that Lorentz-violating wormhole spacetimes behave as effective dispersive optical media, where curvature and symmetry breaking jointly shape wave propagation, scattering, and confinement. This geometric-optical viewpoint provides a unified language linking horizons, turning points, and refractive-index structure, and offers a powerful framework for studying resonances, quasi-normal modes, and, in principle, observational signatures of Lorentz-violating gravity. Future extensions to massive fields, vector and tensor perturbations, and rotating backgrounds may reveal further optical analogies and provide new avenues for probing fundamental spacetime symmetries through wave phenomena.

\end{document}